\documentclass[twoside,12pt]{article}
\usepackage{epsfig, subfig,braket,booktabs}

\usepackage[english]{babel}
\usepackage{amssymb,amsmath,indentfirst}

\newcommand{\pslash}{ \mkern-6mu \not \mkern-4mu p }

\begin{document}

\title{ \vspace{1cm} Chiral nuclear thermodynamics\thanks{Work supported in part by BMBF, GSI and the DFG Cluster of Excellence ``Origin and Structure of the Universe".}}
\author{S.\ Fiorilla, N.\ Kaiser, W. Weise$^*$\\
\\
$^*$Physik-Department, Technische Universit\"{a}t M\"{u}nchen, Garching, Germany}
\maketitle

\begin{abstract}

We calculate the equation of state of nuclear matter for arbitrary isospin-asymmetry up to three loop order in the free energy density in the framework of in-medium chiral perturbation theory. In our approach $ 1\pi$- and $ 2\pi$-exchange dynamics with the inclusion of the $ \Delta$-isobar excitation as an explicit degree of freedom, corresponding to the long- and intermediate-range correlations, are treated explicitly. Few contact terms fixed to reproduce selected known properties of nuclear matter encode the short-distance physics. Two-body as well as three-body forces are systematically included. We find a critical temperature of about 15 MeV for symmetric nuclear matter. We investigate the dependence of the liquid-gas first-order phase transition on isospin-asymmetry. In the same chiral framework we calculate the chiral condensate of isospin-symmetric nuclear matter at finite temperatures. The contribution of the $ \Delta$-isobar excitation is essential for stabilizing the condensate. As a result, we find no indication of a chiral phase transition for densities $ \rho \lesssim 2 \rho_0 $ and at temperatures $ T \lesssim 100 $ MeV. 

\end{abstract}

\section{Introduction}

Nuclear matter is an idealized infinite system of interacting protons and neutrons. It is the low density and low temperature phase of QCD. In this range of densities and temperatures QCD theory is non-perturbative and the active degrees of freedom are no longer the quarks and the gluons, but color singlet states, the hadrons. For the description of nuclear matter a variety of approaches has been developed during the years, being an important issue for its applications to heavy-ion collisions \cite{natowitz} and astrophysics \cite{demorest}. A powerful tool for this task is represented by Chiral Perturbation Theory, an effective field theory that incorporates spontaneous and explicit chiral symmetry breaking. 

The basic idea of our approach relies on the separation between the long-range correlations and the short-distance physics \cite{kaiser1,kaiser2,kaiser3}. While the long- and intermediate-range dynamics in the nuclear medium, described by $ 1\pi$- and $ 2\pi$-exchange with the inclusion of the $ \Delta$-isobar excitation as an explicit degree of freedom, is treated explicitly, the unresolved short-distance physics is encoded in contact terms fixed to reproduce selected known bulk properties of nuclear matter. For example, the contact term associated to the equation of state of isospin-symmetric nuclear matter is fixed imposing that the energy minimum at $ T = 0 $ is $ -16 $ MeV. The correct saturation density, $ \rho_0 \simeq 0.157\,\text{fm}^{-3} $, follows as prediction. For generalization to isospin-asymmetric nuclear matter one has to introduce two more contact terms that are fixed imposing that the asymmetry energy at the saturation point is $ A(\rho_0) \simeq 34 $ MeV \cite{kaiser3}. Many other ground state and single-particle properties and nuclear thermodynamics emerge as predictions. 

Explicit $ 1\pi$- and $ 2\pi$-exchange dynamics are so far treated up to three-loop order in the free energy density. Consider symmetric nuclear matter first. The basic ingredient to perform calculations is the in-medium nucleon propagator. At $ T = 0 $, the nucleon propagator in momentum space has the following representation \cite{kaiser1}:
\begin{equation}\label{propagator}
S_N(p) = \left(\, \pslash + M_N \right) \left[ \frac{i}{p^2- M_N^2 + i\epsilon} - 2 \pi \delta (p^2-M^2) \theta (k_F - |\mathbf{p}|)  \theta(p_0) \right]  \ ,
\end{equation}
where $ k_F $ is the Fermi momentum of nucleons and $ M_N = 939 $ MeV is the nucleon mass. The first term is the known free nucleon propagator, the second term is called medium insertion and takes into account the presence of a filled Fermi sea of nucleons by means of the theta step function. The calculation is then organized according to the number of medium insertions in the diagrams. The relevant many body dynamics come from diagrams with at least two medium insertions. Two-body forces arise from diagrams with two medium insertions;  diagrams with three medium insertions generate the important three body forces. Convergence in the series  of medium insertions is realized for $ k_F \ll \Lambda_\chi \sim 1 $ GeV as long as four-nucleon forces are not relevant. 

\begin{figure}[tbp]
\center
 \subfloat[][Isospin-symmetric nuclear matter.]
 {\includegraphics[width=0.495\textwidth]{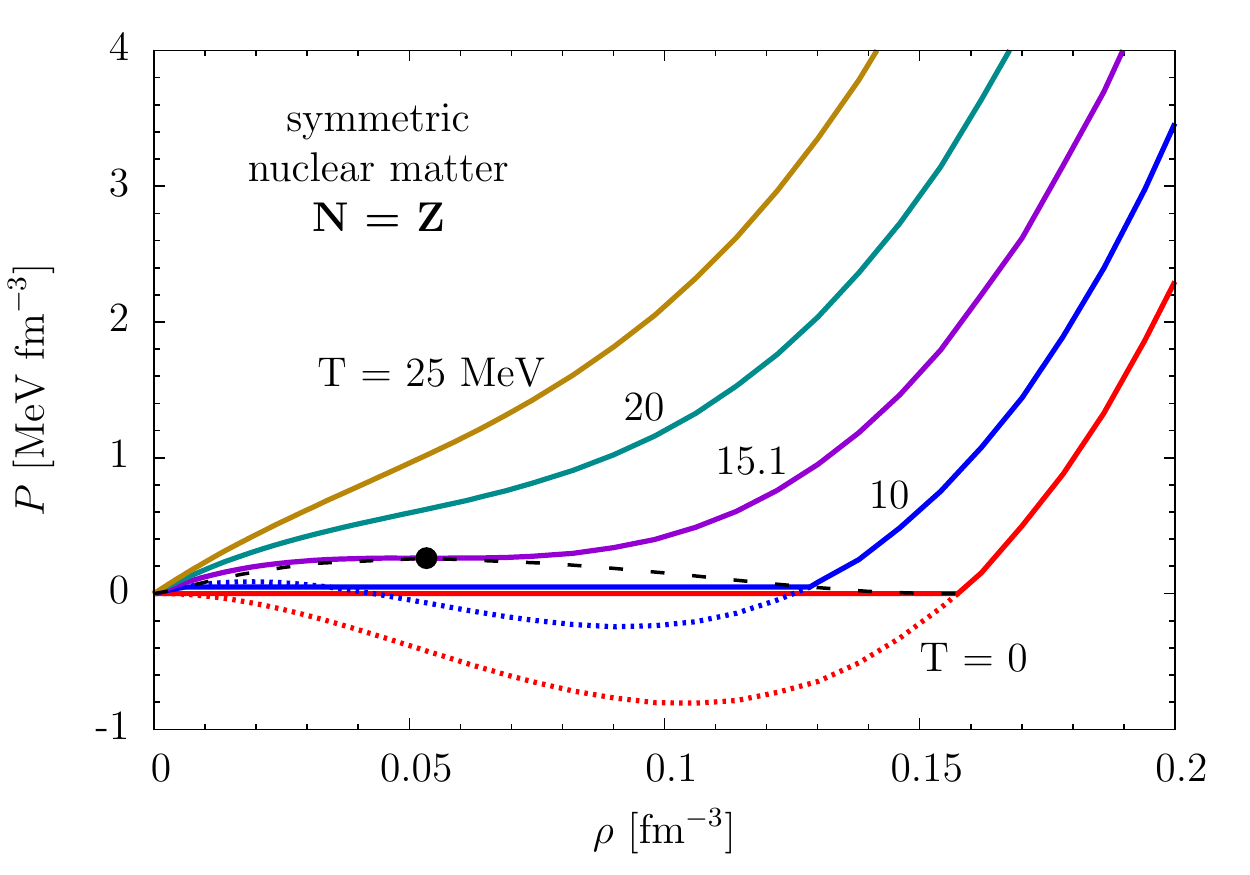}} \ 
  \subfloat[][Neutron-rich matter with $ x_p = 0.1 $.]
 {\includegraphics[width=0.495\textwidth]{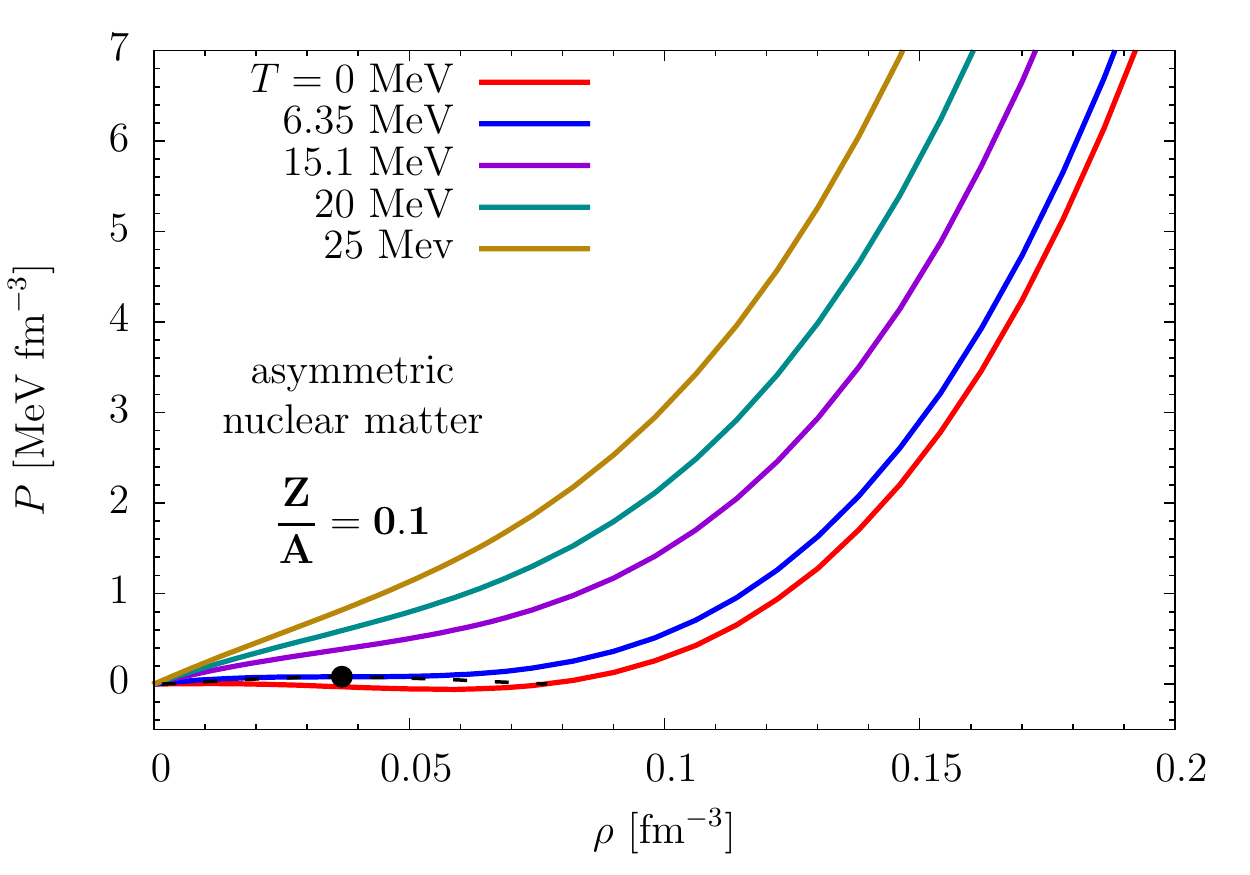}} 
   \caption{Pressure isotherms as a function of nucleon density for symmetric nuclear matter (a) and neutron-rich matter (b). They display a first-order liquid-gas phase transition. The dotted lines at low temperature in (a) show the non-physical behaviour of the equation of state in the phase transition region. The physical pressure is calculated using the Maxwell construction. The dashed lines delimit the boundary of the coexistence region. The dots indicate the critical temperature.} 
 \label{eos}
 \end{figure}

The free energy is written as a sum of convolution integrals \cite{kaiser2}:
\begin{multline} \label{convolution}
	\rho \, \bar{F}(\rho, T) = 4 \int\limits_0^\infty \mbox{d}p \, p \,\mathcal{K}_1(p) \, d(p) + \int\limits_0^\infty \mbox{d}p_1 \int\limits_0^\infty \mbox{d}p_2 \, \mathcal{K}_2(p_1,p_2) \, d(p_1) \, d(p_2) \\
	+ \int\limits_0^\infty \mbox{d}p_1 \int\limits_0^\infty \mbox{d}p_2 \int\limits_0^\infty \mbox{d}p_3 \, \mathcal{K}_3(p_1,p_2,p_3) \, d(p_1) \, d(p_2) \, d(p_3) + \rho \, \mathcal{\bar{A}}(\rho,T) \ ,
\end{multline}
where $  \bar{F}(\rho, T) $ is the free energy per nucleon, $ \mathcal{K}_1 $, $  \mathcal{K}_2 $, $  \mathcal{K}_3 $ are respectively one-body, two-body and three-body kernels, and
\begin{equation}\label{density}
	d(p) = \frac{p}{2\,\pi^2} \left[ 1 + \exp{\frac{p^2/2 M_N - \tilde{\mu}}{T}} \right]^{-1} \ .
\end{equation}
$ \tilde{\mu} $ is the ``one-body'' chemical potential defined through the relation
\begin{equation}
	\rho = 4 \int\limits_0^\infty \mbox{d}p \, p \, d(p) \ .
\end{equation}
$ \mathcal{\bar{A}}(\rho,T) $ is the so-called anomalous contribution and vanishes at $ T = 0 $ \cite{kaiser2}.

The extension of this chiral three loop calculation to isospin-asymmetric nuclear matter requires to take into account the different Fermi seas of protons and neutrons. Consequently we distinguish in Eq.~\eqref{propagator} between proton medium insertion and neutron medium insertion, as well as in Eq.~\eqref{convolution} between proton and neutron distribution functions. Each diagram now involves the sum of all possible combinations of proton and neutron medium insertions with their specific isospin factors.

In section \ref{section2} we present and discuss the equation of state of nuclear matter for different temperatures, featuring the first-order liquid-gas phase transition, and the corresponding phase diagram. In section \ref{section3} we use the scheme above exposed to calculate the chiral condensate of symmetric nuclear matter.

\section{Phase diagram of nuclear matter}\label{section2}
 
  \begin{figure}[tbp]
\centering
\includegraphics[width=.5\columnwidth]{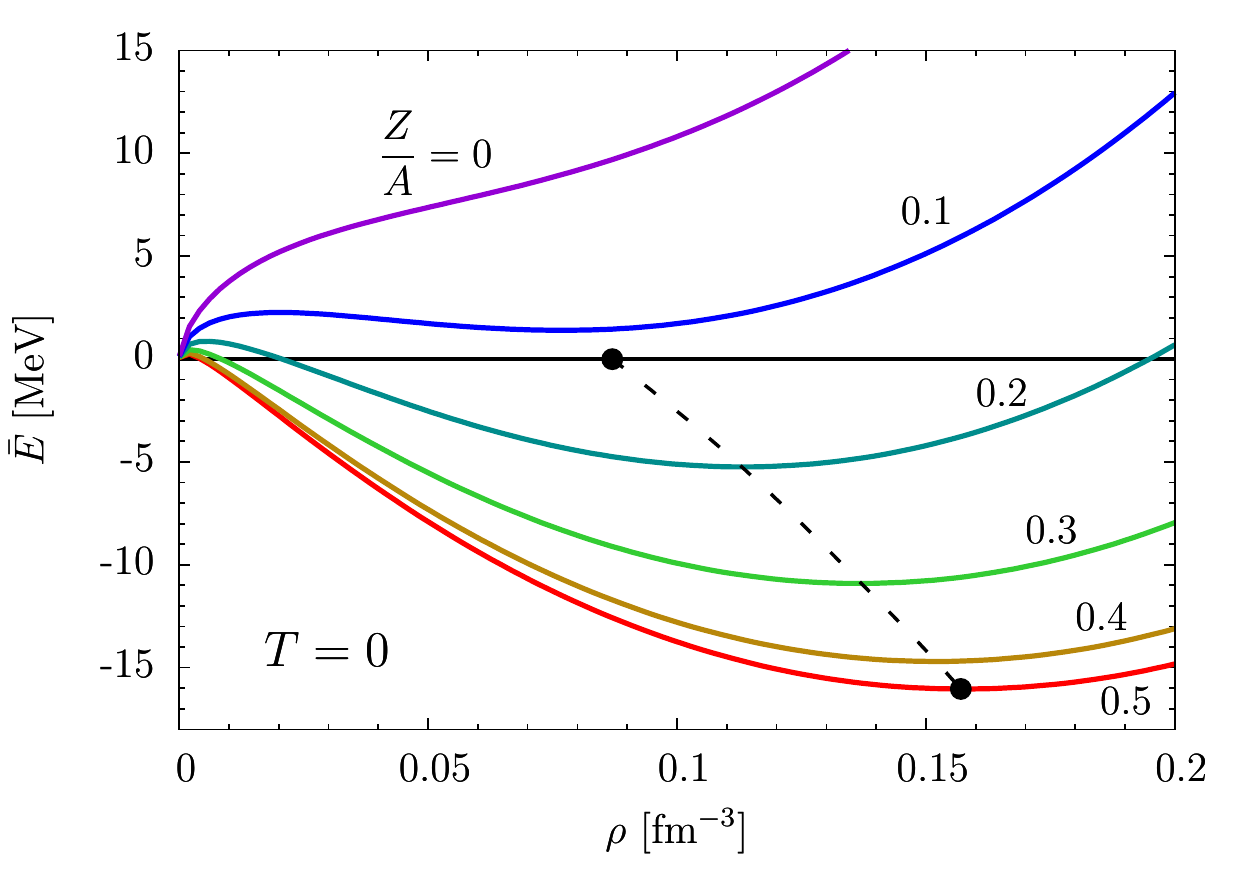}
\caption{Energy per particle  as a function of nucleon density for different proton fractions $ x_p = Z/A $ at $ T = 0$. The dashed line shows the trajectory of the saturation point as $ x_p $ varies. For $ x_p \lesssim 0.12 $ the energy is always positive.}
\label{sat_point}
\end{figure}
 
 In Fig.~\eqref{eos} we show the equation of state of nuclear matter as a function of nucleon density $ \rho $ for a series of temperatures running from 0 up to 25 MeV. The plot (a) displays the pressure isotherms for isospin-symmetric nuclear matter, while the plot (b) is indicative for neutron-rich matter and illustrates the pressure isotherms of nuclear matter with proton fraction $ x_p = 0.1 $ ($ x_p = Z/A $). The qualitative behaviour of these curves is reminiscent of the van der Waals equation of state with its characteristic liquid-gas first-order phase transition. The dotted lines in Fig. (a)  indicate the non-physical behaviour of the equation of state in the liquid-gas coexistence region. This part of the curve is replaced by the physical one (solid line) using the Maxwell construction. The dashed line sets the boundary of the liquid-gas coexistence region, which terminates at the critical point indicated by the dot. For symmetric nuclear matter we find as critical temperature $ T_c \simeq 15.1 $ MeV. With increasing isospin-asymmetry, energy and pressure of nuclear matter at a given density grow in comparison to the symmetric case. At the same time the coexistence region shrinks. In plot (b) one can observe that the critical temperature has diminished to about 6.3 MeV.
 
In Fig.~\eqref{sat_point} we study the dependence of the saturation point on the isospin-asymmetry. The saturation point is defined as the energy minimum at $ T = 0 $. The plot shows the energy per particle of nuclear matter at zero temperature for decreasing proton fraction. The saturation point shifts toward smaller nucleon densities. The corresponding binding energy reduces with continuity, as shown by the dashed line, and vanishes at $ x_p \simeq 0.12 $. Below this value neutron-rich matter is unbound for any density and temperature.
 
 \begin{figure}[htbp]
\centering
 \includegraphics[width=.5\columnwidth]{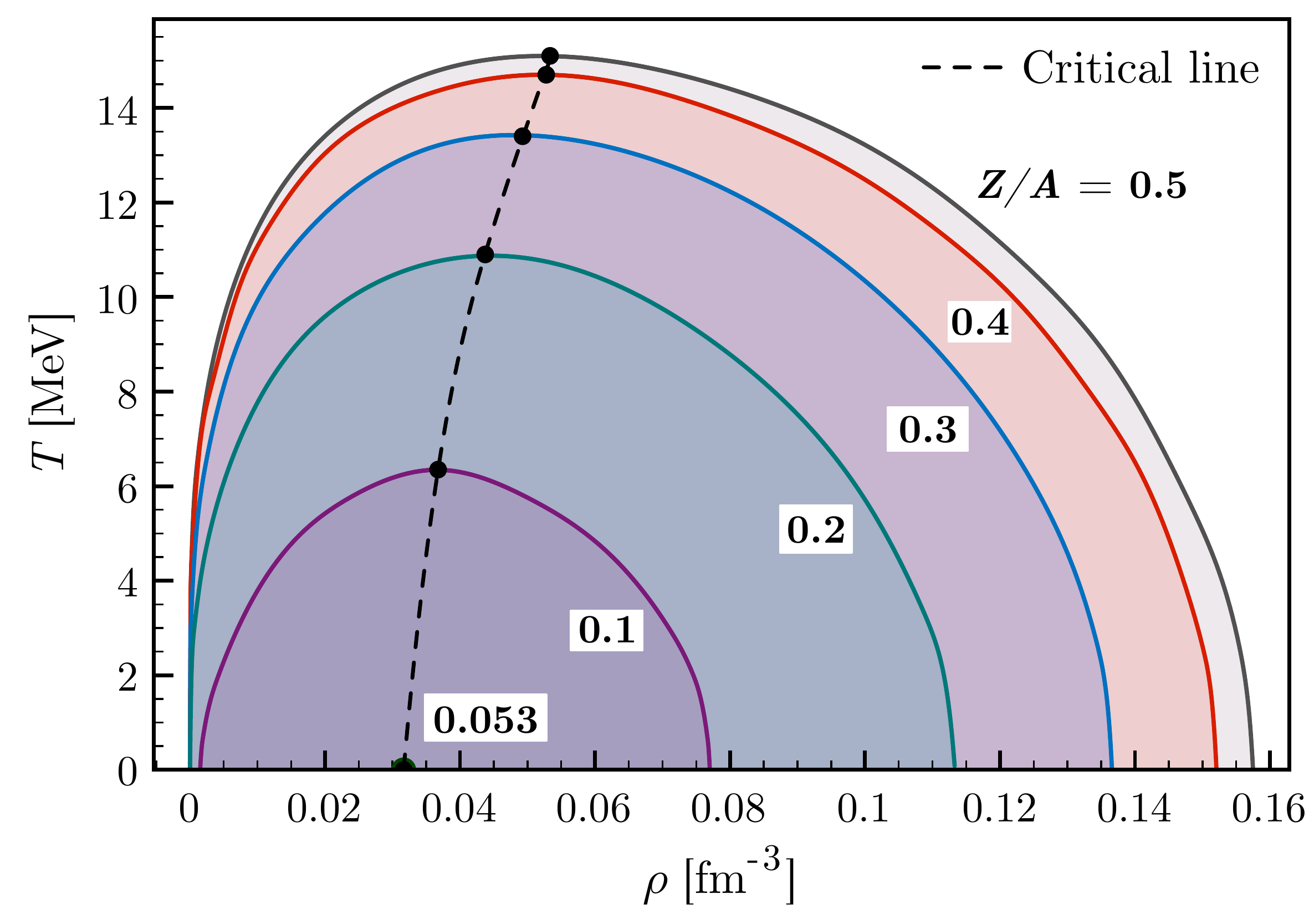}
\caption{$ T-\rho $ phase diagram of nuclear matter for different proton fractions. The dashed line shows  the evolution of the critical point.}
\label{phase_asy}
\end{figure}

The thermodynamic information is well summarized by the $ T - \rho $ (temperature versus nucleon density) phase diagram in Fig.~\eqref{phase_asy}. It visualizes clearly how the liquid-gas coexistence region shrinks with increasing isospin-asymmetry and disappears at $ x_p \simeq 0.05 $. Neutron-rich matter with $ x_p \lesssim 0.05 $ is an interacting Fermi gas and cannot undergo any phase transition. 

 \section{Chiral condensate} \label{section3}

The chiral condensate $ \Braket{\bar{q}q} $ is the order parameter of the chiral phase transition. The non-vanishing of the condensate is connected to the spontaneous breaking of the chiral symmetry of the QCD lagrangian with light quarks. It diminishes with increasing temperature and density. The study of the thermodynamics of the chiral condensate is a key issue in order to locate the boundary of the chiral phase transition in the QCD phase diagram that takes place when the condensate vanishes. The melting of the condensate at high temperature and density restores the chiral symmetry and determines the crossover from the Nambu-Goldstone phase to the Wigner-Weyl realization of the symmetry.

Our starting point is the Hellman-Feynman theorem, which allows to relate the in-medium chiral condensate to the derivative with respect to the quark mass of the free energy per particle of nuclear matter. Using the Gell-Mann-Oakes-Renner relation $ (m_\pi f_\pi) ^2 = -m_q \Braket{0 | \bar{q}q | 0} $, one can rewrite the quark mass derivative as a pion mass derivative, finding the following formula \cite{fio}:
\begin{equation}\label{cond}
\frac{\Braket{\bar{q}q}(\rho,T)}{ \Braket{0|\bar{q}q|0}} = 1 - \frac{\rho}{f_\pi^2} \frac{\partial \bar{F}(\rho,T)}{\partial m_\pi^2} \ .
\end{equation}
The quantities $ f_\pi $ and $ \Braket{0|\bar{q}q|0} $ are taken in the chiral limit ($ m_q, m_\pi \rightarrow 0 $).
The term on the left is the ratio between the in-medium condensate and the vacuum condensate, $ \bar{F} $ is the free energy per particle of nuclear matter. In our model the pion mass appears as an explicit parameter of the free energy and it is possible to calculate its pion mass derivative in a direct way. In the one-body kernel $  \mathcal{K}_1 $ the pion mass dependence is implicit via its dependence on the nucleon mass. The operation of derivation leads to the appearance of the pion nucleon sigma term $ \sigma_N = m_\pi^2\, \partial M/\partial m_\pi^2 $. For our calculation we choose the empirical value $ \sigma_N = (45 \pm 8) $ MeV \cite{sigma}. At $ T = 0 $ the chiral condensate assumes a simpler representation \cite{kaiser7}:
\begin{equation}
\frac{\Braket{\bar{q}q}(\rho)}{ \Braket{0|\bar{q}q|0}} = 1 - \frac{\rho}{f_\pi^2} \left\{ \frac{\sigma_N}{m_\pi^2} \left( 1 - \frac{3\,k_F^2}{10 M^2} \right) + \frac{\partial \bar{E}_{int}(k_F)}{\partial m_\pi^2} \right\} \ .
\end{equation}
The term in the round bracket is the contribution of the non-interacting Fermi gas. It is essentially proportional to the nucleon density and is the leading term of the chiral condensate. The second term is the contribution of the interaction driven by the pion exchange dynamics to the condensate ($ \bar{E}_{int} $ is the interaction energy per particle).

 \begin{figure}[htbp]
\centering
 \includegraphics[width=.5\columnwidth]{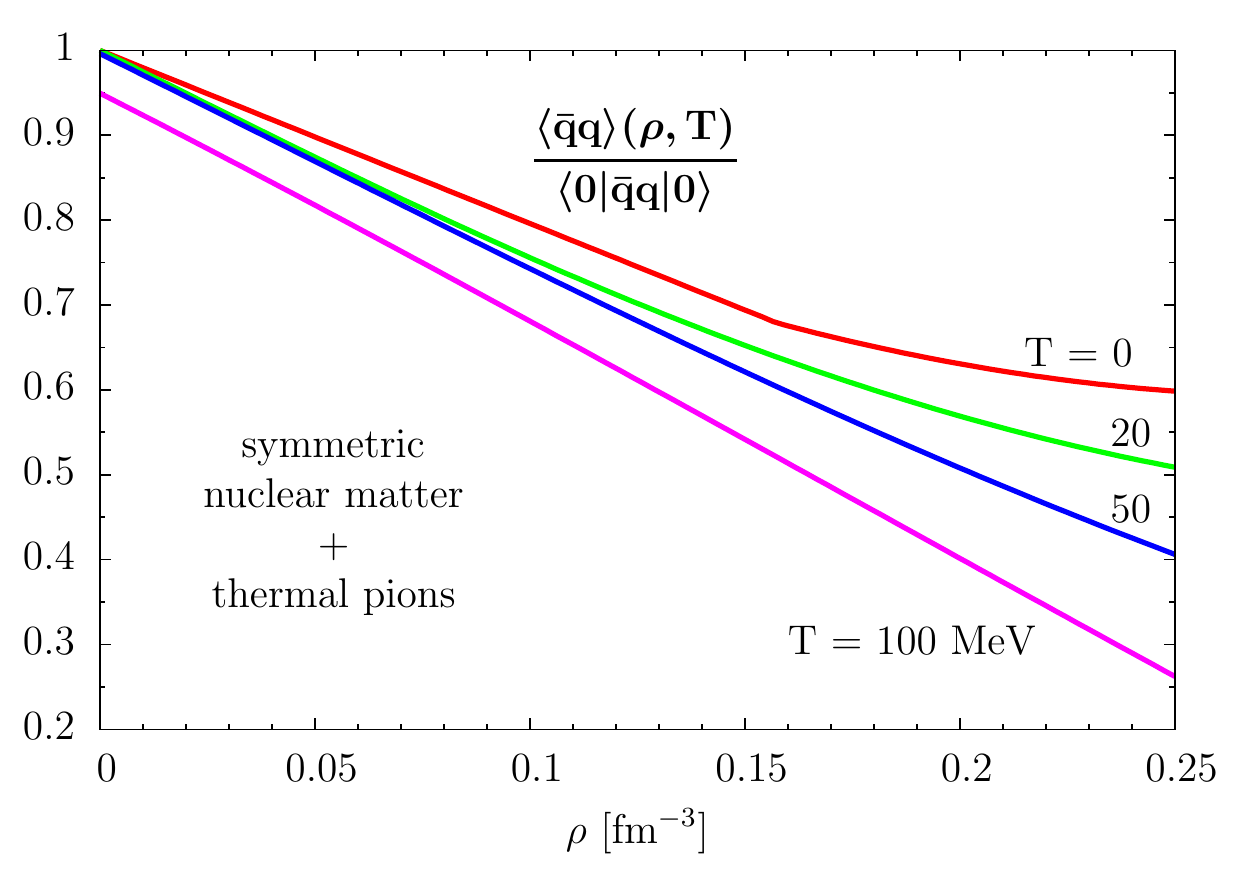}
\caption{Ratio of the chiral condensate of symmetric nuclear matter to its vacuum value as a function of the nucleon density $ \rho $ for different temperatures up to 100 MeV. The contribution of the thermal pions is included. In the liquid-gas coexistence region at low temperatures ($ T \lesssim 15 $ MeV) the quark condensate decreases linearly.}
\label{condensate}
\end{figure}

In Fig.~\ref{condensate} we show the behaviour of the condensate ratio \eqref{cond} as a function of nucleon density for different temperatures from 0 up to $ 100 $ MeV. The contribution of the thermal pions is also included \cite{kaiser}, consisting in a further reduction of the temperature dependent condensate. The effect of the thermal pions is appreciable at high temperatures, $ T > 50 $ MeV. At low temperature ( $ T \lesssim 15 $ MeV) one can recognize the trace of the liquid-gas first-order phase transition on the condensate curves. Using the Maxwell construction one finds that the chiral condensate at a given temperature decreases linearly in the liquid-gas coexistence region (see curve at $ T = 0 $). The net effect of the interactions is to counteract the reduction due to the linear density term. A crucial role is played by the $ \Delta$-isobar excitation. Its contribution is large in comparison to that of the other terms and stabilizes the condensate. The bending of the curves at large densities in Fig.~\ref{condensate} is a consequence of the inclusion of the $ \Delta$-isobar excitation as an explicit degree of freedom. Had we taken into account only the pion exchange dynamics without the $ \Delta$-isobar excitation, the system would have reached chiral restoration not far from the saturation point. With increasing temperature the effect of the interaction weakens and the linear behaviour is practically restored. This recovery is the result of a balance between attractive and repulsive correlations and their mass dependence. 

From systematics in the variation of the condensate with nucleon density and temperature presented in Fig.~\ref{condensate} no rapid tendency toward chiral symmetry restoration is seen, at least for densities up to $ \rho \lesssim 2\,\rho_0 $  and temperatures $ T \lesssim 100 $ MeV. Being far from the chiral phase transition, our chiral perturbative approach is reliable in this range of densities and temperatures.

\section{Summary}

In the present review we have presented recent calculations of the equation of state of nuclear matter \cite{fio2} and of the chiral condensate \cite{fio} of isospin-symmetric nuclear matter in the framework of in-medium chiral perturbation theory up to three-loop order in the free energy density. 

In our approach long- and intermediate-distance correlations driven by pion exchange dynamics are treated explicitly, while the unresolved short-distance interaction is taken into account by fixing a few contact terms in order to reproduce some selected bulk properties of nuclear matter. Many other ground state, single-particle and thermodynamic properties follow as a prediction of the model. Correlations in the nuclear medium include two- and three-body forces. The resulting equation of state features the characteristic first-order liquid-gas phase transition of nuclear matter, whose dependence on isospin-asymmetry is systematically investigated. We find for isospin-symmetric nuclear matter a critical temperature of about 15 MeV.

The chiral condensate of isospin-symmetric nuclear matter is calculated by differentiating the free energy per particle with respect to the pion mass, which appears as an explicit parameter of the model. Correlations due to $ \Delta$-isobar excitation play the important role of stabilizing the condensate. As a result, we find no indication of a chiral phase restoration for densities up to about twice the density of normal nuclear matter and at temperatures below 100 MeV. This is the first calculation of the chiral condensate at finite density and temperature that systematically includes in-medium two-pion exchange interactions. 

The purpose of the present work is to set realistic nuclear physics constraints for ongoing discussions of the QCD phase diagram.

\end{document}